\documentclass[runningheads]{llncs}
\usepackage{multirow}

\usepackage[T1]{fontenc}
\usepackage{graphicx}
\usepackage{booktabs}
\usepackage{marvosym}
\begin{document}
\title{EchoVoices: Preserving Generational Voices and Memories for Seniors and Children}
\titlerunning{EchoVoices}
%
\author{Haiying Xu\inst{1}\textsuperscript{*} \orcidID{0009-0004-1965-8416} \and Haoze Liu\inst{1}\textsuperscript{*} \orcidID{0009-0008-5775-984X} \and Mingshi Li\inst{1} \orcidID{0009-0009-4213-5409} \and Siyu Cai\inst{1}  \orcidID{0009-0002-5687-8149} \and Guangxuan Zheng\inst{1}  \orcidID{0009-0007-2918-2855} \and Yuhang Jia\inst{1} \orcidID{0009-0001-2407-0789} \and Jinghua Zhao\inst{1} \orcidID{0009-0008-9648-7373} \and Yong Qin\inst{1}\orcidID{0009-0000-2748-3020}\textsuperscript{$^{\dagger}$}} 
\authorrunning{H. Xu, H. Liu et al.}
%
\institute{College of Computer Science, Nankai University\\[0.5em]
\textbf{Correspondence:} 2212180@mail.nankai.edu.cn, qinyong@nankai.edu.cn
}
\maketitle
          
%
\def\thefootnote{*}\footnotetext{These authors contributed equally to this work, sorted by surname.}
\def\thefootnote{$\dagger$}\footnotetext{Corresponding author.}
\def\thefootnote{\arabic{footnote}}

\vspace{-1em}

\vspace{-0.5em}

\begin{abstract}


Recent breakthroughs in intelligent speech and digital human technologies have primarily targeted mainstream adult users, often overlooking the distinct vocal patterns and interaction styles of seniors and children. These demographics possess distinct vocal characteristics, linguistic styles, and interaction patterns that challenge conventional ASR, TTS, and LLM systems. To address this, we introduce EchoVoices, an end-to-end digital human pipeline dedicated to creating persistent digital personas for seniors and children, ensuring their voices and memories are preserved for future generations. Our system integrates three core innovations: a k-NN-enhanced Whisper model for robust speech recognition of atypical speech; an age-adaptive VITS model for high-fidelity, speaker-aware speech synthesis; and an LLM-driven agent that automatically generates persona cards and leverages a RAG-based memory system for conversational consistency. Our experiments, conducted on the SeniorTalk and ChildMandarin datasets, demonstrate significant improvements in recognition accuracy, synthesis quality, and speaker similarity. EchoVoices provides a comprehensive framework for preserving generational voices, offering a new means of intergenerational connection and the creation of lasting digital legacies.\footnote{Our project is available at: \url{https://github.com/lhz191/EchoVoices.git}}



\end{abstract}
\section{Introduction}

\vspace{-0.5em}



Recent advances in intelligent speech and digital human technologies have produced sophisticated conversational agents for mainstream adult users. These systems, however, often fail to accommodate the unique needs of seniors and children. Due to distinct vocal patterns, prosody, and linguistic styles inherent to these age groups, conventional Automatic Speech Recognition (ASR), Text-to-Speech (TTS), and Large Language Model (LLM) frameworks struggle with accuracy and naturalness, limiting their accessibility and utility.

This technological gap carries a significant cost: the rich oral histories of the elderly and the critical linguistic development of the young risk being unrecorded in our digital age. To address this, we propose EchoVoices, a system founded on the principle of holistic preservation. Rather than pursuing simple technical fixes, our work aims to create persistent and authentic digital personas that capture the full essence of an individual's identity, including their memories, personality, and unique way of speaking.

The EchoVoices pipeline integrates specialized modules for age-adaptive speech recognition and synthesis with an LLM-driven agent for consistent memory management. By combining these components, we create expressive digital avatars capable of meaningful, context-aware interaction. This approach provides a novel framework for preserving generational wisdom and creating lasting digital legacies, offering a new technological medium for intergenerational connection.



Our main contributions are as follows:
\vspace{-0.5em}
\begin{itemize}

    \item We developed \textbf{EchoVoices}, a holistic framework to generate expressive digital personas for seniors and children, which integrates specialized ASR, TTS, and retrieval-augmented LLM modules to authentically capture and preserve their unique voices, memories, and interaction styles.
      

   \item We introduce a \textbf{k-NN} augmentation for Whisper that significantly improves ASR on atypical speech, and a \textbf{two-stage training strategy} for VITS that yields high-fidelity, age-adaptive speech synthesis.


    \item We deploy an LLM-driven agent that automatically distills a user's \textbf{persona card} from conversation and uses a RAG-based memory to maintain a consistent and evolving digital identity.

\end{itemize}

\vspace{-0.5em}

\vspace{-0.5em}
\vspace{-0.5em}

\section{Related Work}
\vspace{-0.5em}

\vspace{-0.5em}

\subsection{Speech Recognition and Synthesis}
\vspace{-0.5em}
Large-scale ASR models like Whisper~\cite{whisper} show impressive general capabilities but often falter on the atypical speech of seniors and children. The k-Nearest-Neighbor (k-NN) paradigm, which augments model predictions by retrieving similar examples from a datastore~\cite{knn-lm}, offers a powerful method for domain adaptation without costly retraining. While k-NN has been applied to CTC-based ASR~\cite{zhou2023knnctc}, its use to enhance large encoder-decoder models like Whisper for the distinct challenges of elderly and child speech remains underexplored. For speech synthesis, while models like VITS~\cite{vits} enable high-quality, multi-speaker generation, they require extensive resources for zero-shot adaptation~\cite{valle}. Our work addresses these gaps by adapting the k-NN framework to improve fine-tuned Whisper models and employing a resource-efficient, two-stage training strategy for VITS, specifically for these demographics.

\vspace{-0.5em}
\vspace{-0.5em}

\subsection{LLM-Driven Digital Personas}
\vspace{-0.5em}
The paradigm of Large Language Models (LLMs) now extends to powering sophisticated, autonomous agents, where maintaining a consistent persona is critical. Retrieval-Augmented Generation (RAG)~\cite{rag} is a key methodology for this, grounding LLMs in external knowledge to ensure factual consistency. Our work applies this by first using an LLM to distill a foundational 'persona card' from dialogue and then implementing a RAG-based memory system to ensure a coherent and evolving digital identity. This is situated within the broader field of digital human generation, where ASR is fundamental for animating talking faces using methods like Wav2Lip~\cite{wav2lip} or GeneFace~\cite{geneface}. Our work contributes to this field by enhancing ASR for seniors and children, enabling more empathetic and effective applications.

\vspace{-0.5em}
\vspace{-0.5em}

\vspace{-0.5em}
\section{Method}
\vspace{-0.5em}

\subsection{System Overview}



\begin{figure}
    \centering
    \includegraphics[width=1\linewidth]{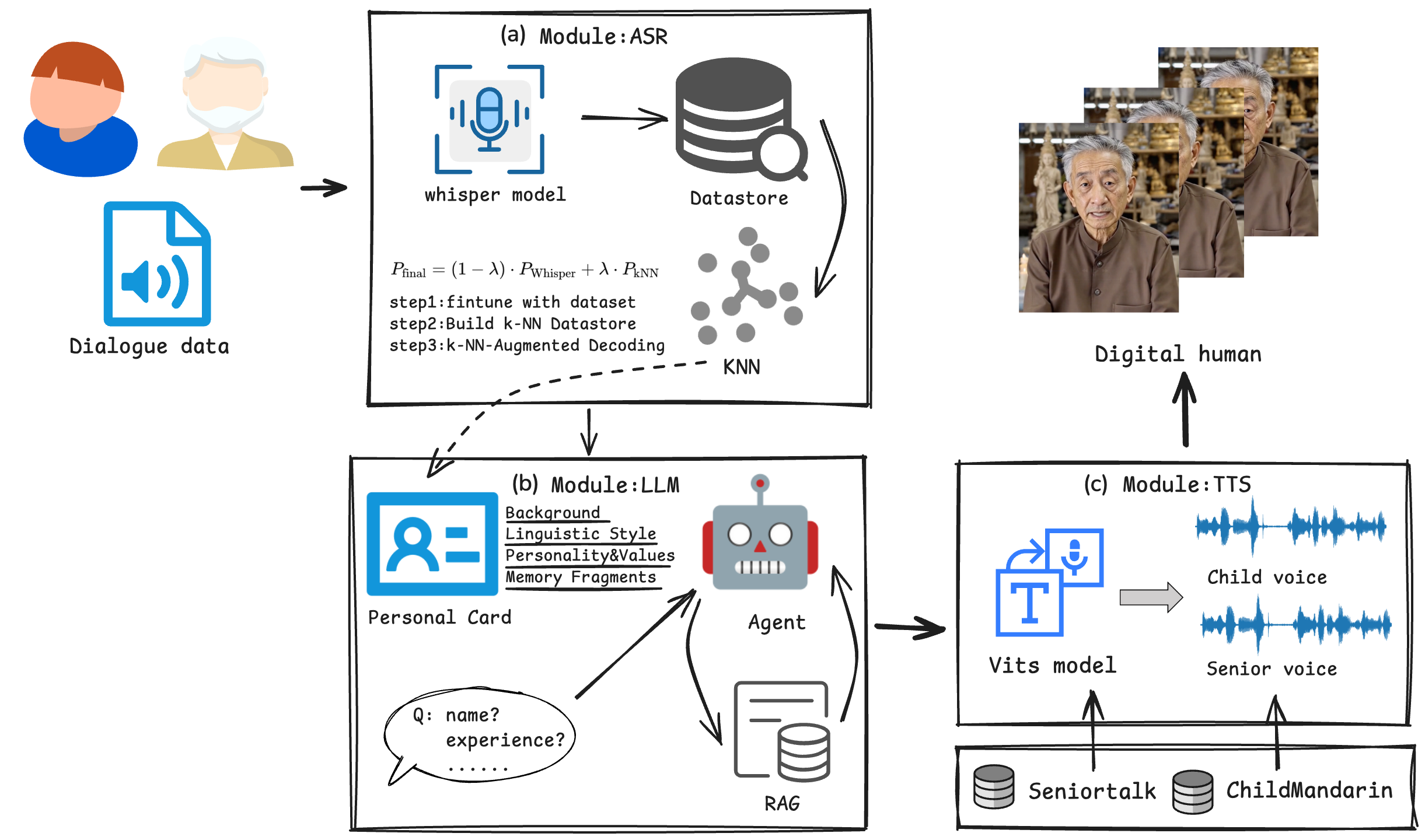}
    \caption{The EchoVoices System Pipeline. (a) A spoken query from a senior or child is transcribed by our k-NN enhanced Whisper ASR model. (b) The text is processed by an LLM-driven agent, which uses RAG to query a memory database and generate a persona-consistent response. (c) The response text is synthesized into age-appropriate speech by a two-stage fine-tuned VITS model.}
    \label{fig:enter-label2}
\end{figure}

Our  system enables personalized, cross-age speech interaction by integrating speech recognition, language understanding, speech synthesis, and talking-face generation in a single pipeline (Fig.~\ref{fig:enter-label2}). The input is a spoken query from a child or elderly speaker, and the output is a synchronized talking-face video with high-quality, speaker-aware synthetic speech.

We  employ a fine-tuned Whisper model to transcribe the input audio into text, adapting it to elderly and child speech domains. The recognized text is then passed to a large language model (LLM) equipped with retrieval-augmented generation (RAG) and Self-Talk prompting, which generates a context-aware, informative response.
The LLM output is then fed into a two-stage trained VITS model, which synthesizes personalized speech for elderly or child voice characteristics. Finally, we use Wav2Lip combined with GFPGAN to render a synchronized, photorealistic talking-face video driven by the generated speech.

This modular design enables both speaker-specific adaptation and cross-modal generation, supporting engaging and inclusive voice interaction for underrepresented user groups.

\vspace{-0.5em}

\vspace{-1em}

\subsection{k-NN Enhanced ASR for Specialized Demographics}
To augment our fine-tuned Whisper model, we integrate a k-Nearest-Neighbors (k-NN) mechanism that leverages instance-based knowledge at inference time. First, we construct a domain-specific datastore by processing the training set (e.g., elderly or child speech) with the fine-tuned model. For each token, the final hidden state from the Whisper decoder is extracted as the key, and the corresponding ground-truth token is the value. These key-value pairs are indexed in a Faiss~\cite{faiss} datastore for efficient retrieval. During inference, at each step of the beam search, the decoder's current hidden state is used to query the datastore for the k-nearest neighbors. Their corresponding tokens are used to form a non-parametric probability distribution P\textsubscript{kNN}, which is interpolated with the model's original distribution P\textsubscript{Whisper} to produce the final prediction:
\[ P_{final} = (1 - \lambda) P_{Whisper} + \lambda P_{kNN} \]
where $\lambda$ is a hyperparameter balancing the two distributions. This allows the model to correct predictions for domain-specific or rare patterns by consulting an explicit memory.

\vspace{-0.5em}

\vspace{-0.5em}

\subsection{Age-Adaptive TTS with VITS Fine-tuning}
\vspace{-0.5em}
To synthesize age-appropriate speech, we employ a two-stage training strategy on the VITS architecture~\cite{vits}. First, a model is pretrained on a large subset (90\%) of the target demographic data (SeniorTalk or ChildMandarin) to learn its general acoustic and prosodic characteristics, such as vocal tremor in seniors or higher fundamental frequencies in children. For adaptation to new, unseen speakers, we then re-initialize the speaker embedding layer with \( \mathbf{e}_{speaker}^{(0)} \sim \mathcal{N}(0, \sigma^2 \mathbf{I}) \) and fine-tune the model on the held-out 10\% of speakers. This re-initialization prevents overfitting to the pretrained identities and facilitates rapid adaptation to the new speaker's unique vocal characteristics, ensuring natural and demographically appropriate speech synthesis.

\vspace{-1em}

\subsection{LLM-Driven Persona Generation}

\vspace{0.5em}

\begin{figure}
    \centering
    \includegraphics[width=1\linewidth]{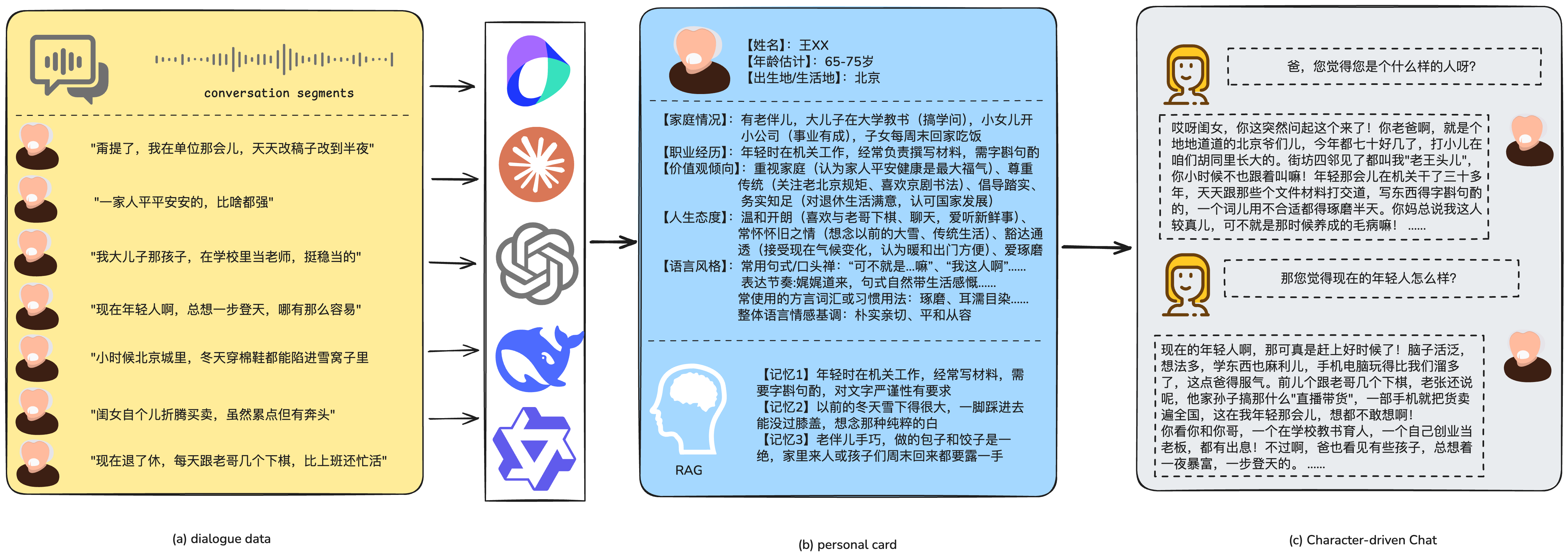}
     \caption{The LLM-driven agent pipeline. (a) Spoken dialogue is first collected and transcribed into text using the fintuned ASR model. (b) The transcribed text is processed by a large language model, which extracts user-specific identity cards and retrieves relevant memory information using a retrieval-augmented generation (RAG) mechanism. (c) Based on the persona card and retrieved memory, the agent generates character-consistent dialogue responses as illustrative examples.}
    \label{fig:enter-label}
\end{figure}

\vspace{-1.5em}

To imbue the digital human with a consistent and evolving identity, we employ an LLM-driven agent centered around a dynamic persona card. This card, which functions as the agent's foundational prompt, is a structured summary of the user's background, linguistic style, and key memories, automatically distilled and updated by an LLM analyzing the conversation. For conversational continuity, this core persona is augmented by a RAG-based memory system. Salient facts from the dialogue are stored in a vector database and retrieved via similarity search when generating a new response. These retrieved memories are then combined with the persona card in the LLM's prompt, ensuring every response is both in-character and grounded in the history of the interaction.

\vspace{-0.5em}
\vspace{-0.5em}
\vspace{-0.5em}

\section{Experiments}
\vspace{-0.5em}
\subsection{Dataset}
\vspace{-0.5em}
We use two Mandarin speech datasets targeting underrepresented age groups: SeniorTalk~\cite{chen2024seniortalk} for elderly speech and ChildMandarin~\cite{zhou2024childmandarin} for young children.

\textbf{SeniorTalk} contains 55.53 hours of spontaneous speech from 202 speakers aged 75--85, collected across 16 provinces. It is balanced in gender, region, and age, and includes annotations such as speaker ID, timestamps, and accent tags.

\textbf{ChildMandarin} includes 41.25 hours of speech from 397 speakers aged 3--5, recorded via smartphones in 22 provinces, with balanced gender distribution. Each audio is paired with character-level transcriptions and metadata including age, gender, and region.

\vspace{-1em}



\subsection{Experimental Setup}
\subsubsection{ASR Fine-tuning with Whisper}


For the ASR task, we evaluate four sizes of the Whisper model (tiny, base, small, and medium) on both datasets, measuring performance in Character Error Rate (CER). We compare three settings: (1) \textbf{Zero-shot} performance using the off-the-shelf models, (2) \textbf{Fine-tuned} models adapted on each domain-specific dataset, and (3) our proposed \textbf{Fine-tuned + k-NN} approach, which augments the fine-tuned model with k-NN decoding.
\vspace{0.5em}

\subsubsection{Results}
We present the ASR results on the SeniorTalk and ChildMandarin datasets in Table~\ref{tab:senior_results} and Table~\ref{tab:child_results}, respectively. The results are reported in terms of CER (\%).
\vspace{-1em}

\vspace{-0.5em}
\vspace{-0.5em}

\begin{table}[h!]
  \centering
  \caption{CER (\%) on the SeniorTalk dataset. Fine-tuning and k-NN augmentation significantly improve performance.}
  \label{tab:senior_results}
  \begin{tabular}{l c c c c}
    \hline
    \textbf{Model} & \textbf{Parameters} & \textbf{Zero-shot} & \textbf{Fine-tuned} & \textbf{Fine-tuned + k-NN} \\
    \hline
    Whisper-tiny & 39M & 82.85 & 29.73 & \textbf{26.69} \\
    Whisper-base & 74M & 67.54 & 22.70 & \textbf{18.85} \\
    Whisper-small & 244M & 56.34 & 17.59 & \textbf{15.84} \\
    Whisper-medium & 769M & 48.61 & 27.96 & \textbf{14.78} \\
    \hline
  \end{tabular}
\end{table}
\vspace{-0.5em}
\vspace{-0.5em}
\vspace{-0.5em}
\vspace{-0.5em}
\vspace{-0.5em}

\begin{table}[h!]
  \centering
   \caption{CER (\%) on the ChildMandarin dataset. The high CER for the medium fine-tuned model reflects a high deletion rate, a failure mode our k-NN approach mitigates.}
  \label{tab:child_results}
  \begin{tabular}{l c c c c}
    \hline
    \textbf{Model} & \textbf{Parameters} & \textbf{Zero-shot} & \textbf{Fine-tuned} & \textbf{Fine-tuned + k-NN} \\
    \hline
    Whisper-tiny & 39M & 72.49 & \textbf{28.73} & 29.17 \\
    Whisper-base & 74M & 51.10 & \textbf{22.31} & 22.58 \\
    Whisper-small & 244M & 30.93 & 21.03 & \textbf{18.36} \\
    Whisper-medium & 769M & 24.45 & 81.49 & \textbf{17.66} \\
    \hline
  \end{tabular}
\end{table}
\vspace{-0.5em}
\vspace{-0.5em}

\vspace{-0.5em}
\vspace{-0.5em}
\vspace{-0.5em}


The results demonstrate a clear and consistent trend: fine-tuning provides a dramatic improvement over the zero-shot baseline across both datasets, underscoring the necessity of domain adaptation for these specialized demographics. On the SeniorTalk dataset, our k-NN enhancement method consistently pushes the performance further, providing an additional relative improvement across all model sizes and achieving a best overall CER of 14.78\% with the medium model. This highlights the benefit of combining parametric adaptation with non-parametric, instance-based knowledge retrieval.

On the ChildMandarin dataset, our method shows compelling results. Due to insufficient training data, the fine-tuned medium model suffers a critical failure, frequently producing no output and resulting in a prohibitive 81.49\% CER. Our k-NN augmentation not only improves performance but robustly rectifies this failure, reducing the CER to 17.66\%. This result powerfully demonstrates that the k-NN mechanism can serve as a crucial guide for the model, ensuring it produces a valid output where it would otherwise fail.






\vspace{-0.5em}
\vspace{-0.5em}
\vspace{-0.5em}

\subsection{TTS Fine-tuning with VITS}
\vspace{-0.5em}

\subsubsection{Experimental Setup}
We evaluate our TTS approach using a speaker independent protocol on both datasets, partitioning speakers into 90\% for pretraining and 10\% for fine-tuning. We compare direct fine-tuning on the 10\% subset against our proposed method which first pretrains on the 90\% subset. Synthesis quality is measured by the Character Error Rate (CER) from our ASR model on synthesized audio (CER\textsubscript{f}) versus ground-truth audio (CER\textsubscript{g}). Speaker similarity is evaluated using ECAPA-TDNN, x-vector, and pyannote.

\vspace{-1.5em}

\subsubsection{Results}

As shown in Table~\ref{tab:tts_compact_results}, our two-stage training strategy consistently improves speaker similarity across all metrics. For instance, on the SeniorTalk dataset, the ECAPA-TDNN score increases from 0.5608 to 0.5644, and similar gains are observed for ChildMandarin. This indicates that pretraining on a larger, demographically-matched dataset allows the model to establish a robust prior of age-specific acoustic features. This robust prior not only enhances speaker similarity but also enables faster convergence during the fine-tuning stage for new, unseen speakers.

Furthermore, this pretraining significantly enhances the intelligibility of the synthesized speech, as reflected by the reduction in Character Error Rate on the synthesized audio (CER\textsubscript{f}). On the SeniorTalk dataset, CER\textsubscript{f} drops from 50.52\% to 45.63\%, with comparable improvements on ChildMandarin. The effectiveness of this approach stems from the division of labor: the pretraining stage captures general characteristics like vocal tremor in seniors or higher fundamental frequencies in children, while the subsequent fine-tuning stage, aided by speaker embedding re-initialization, can focus more effectively on capturing the unique timbre of the target speaker. This results in more natural and demographically appropriate synthetic speech.

\vspace{-0.5em}
\vspace{-0.5em}
\vspace{-0.5em}

\begin{table}[h!]
  \centering
  \caption{CER and speaker similarity for TTS fine-tuning on SeniorTalk and ChildMandarin datasets. “wi pt” denotes pretraining on 90\% of speakers. Fine-tuning is conducted on the remaining 10\% of unseen speakers.}
  \label{tab:tts_compact_results}

  \begin{tabular}{l ccccc}
    \hline
    \multirow{2}{*}{\textbf{Model}} & 
    \multicolumn{2}{c}{\textbf{CER }} & 
    \multicolumn{3}{c}{\textbf{Speaker Similarity}} \\
    \cline{2-3} \cline{4-6}
    & \textbf{CER\textsubscript{g}} & \textbf{CER\textsubscript{f}} 
    & \textbf{ECAPA} & \textbf{x-vector} & \textbf{pyannote} \\
    \hline
    SeniorTalk (wo pretrain)  & 42.60 & 50.52 & 0.5608 & 0.9503 & 0.5593 \\
    SeniorTalk (wi pretrain  & \textbf{36.11} & \textbf{45.63} & \textbf{0.5644} & \textbf{0.9512} & \textbf{0.5637} \\
    ChildMandarin (wo pretrain) & 55.44 & 42.58 & 0.5999 & 0.9566 & 0.5899 \\
    ChildMandarin (wi pretrain) & \textbf{48.93} & \textbf{39.18} & \textbf{0.6116} & \textbf{0.9583} & \textbf{0.5972} \\
    \hline
  \end{tabular}
\end{table}
\vspace{-0.5em}

\vspace{-0.5em}

\vspace{-2.5em}

\section{Limitation}

\vspace{-0.5em}





While our current TTS system based on VITS effectively improves synthesis quality through speaker-specific fine-tuning, it incurs additional computational cost. Future work may explore zero-shot speaker adaptation techniques to improve scalability and eliminate the need for per-speaker customization.Secondly, though the k-NN enhancement proves effective, it introduces increased inference-time latency and computational overhead, highlighting the need for future optimization strategies such as datastore pruning or vector quantization. Finally, while we propose a cascaded system for elderly and child users, its modular design may introduce latency and inefficiency. Future research may explore the development of a unified end-to-end multimodal large model to improve efficiency and streamline the interaction process.
\vspace{-1.5em}

\section{Conclusion}
\vspace{-0.5em}

In this paper, we presented EchoVoices, a holistic framework designed to create authentic digital personas for seniors and children, addressing the shortcomings of conventional speech technologies for these demographics. We demonstrated the effectiveness of a modular approach that integrates three key innovations: a k-NN augmented Whisper model for robust, domain-adapted ASR; a two-stage training strategy for VITS that yields high-fidelity, age-appropriate synthetic speech; and an LLM-driven agent that maintains a consistent persona through automatically generated persona cards and a RAG-based memory. Our experiments validated this approach, showing not only significant improvements in ASR accuracy—where the k-NN mechanism was crucial in rectifying catastrophic model failure on child speech—but also enhanced speaker similarity and intelligibility in TTS. Ultimately, EchoVoices provides a robust and comprehensive blueprint for developing more inclusive AI, enabling the preservation of generational voices and creating new pathways for digital legacy and intergenerational connection.

\vspace{-1.5em}

\bibliographystyle{unsrt}
\bibliography{main}

\end{document}